\documentclass[reprint]{revtex4-2}
\usepackage[latin9]{inputenc}
\usepackage{array}
\usepackage{amstext}
\usepackage{amsmath}
\usepackage{graphicx}
\usepackage{esint}
\usepackage{cancel}
\usepackage[section]{placeins}
\usepackage[normalem]{ulem}
\usepackage{float}

\usepackage{color}
\usepackage{xcolor}
\usepackage{titlesec}

\usepackage[unicode=true,pdfusetitle,
 bookmarks=true,bookmarksnumbered=false,bookmarksopen=false,
 breaklinks=false,pdfborder={0 0 1},backref=false,colorlinks=false]
 {hyperref}



\begin{document}
\title{Elastic fingering in a rotating Hele-Shaw cell}
\author{Benjamin Foster}
\email{ben\_foster@berkeley.edu}
\address{Physics Department, University of California at Berkeley, Berkeley,
California 94720, USA}
\author{Edgar Knobloch}
\email{knobloch@berkeley.edu}
\address{Physics Department, University of California at Berkeley, Berkeley,
California 94720, USA}

\begin{abstract}

We consider the steady-state fingering instability of an elastic membrane separating two fluids of different density under external pressure in a rotating Hele-Shaw cell.  Both inextensible and highly extensible membranes are considered, and the role of membrane tension is detailed in each case.  Both systems exhibit a centrifugally-driven Rayleigh-Taylor--like instability when the density of the inner fluid exceeds that of the outer one, and this instability competes with the restoring forces arising from curvature and tension, thereby setting the finger scale. Numerical continuation is used to compute not only strongly nonlinear primary finger states up to the point of self-contact but also secondary branches of mixed modes and circumferentially localized folds as a function of the rotation rate and the externally imposed pressure. Both reflection-symmetric and symmetry-broken chiral states are computed. The results are presented in the form of bifurcation diagrams.  The ratio of system scale to the natural length scale is found to determine the ordering of the primary bifurcations from the unperturbed circle state as well as the solution profiles and onset of secondary bifurcations.

\end{abstract}
\noindent{\it Keywords\/}: {\noindent Rayleigh-Taylor instability; nonlinear elastica; wrinkling; bifurcation}

\maketitle

\section{Introduction}

The Saffman-Taylor instability, also known as the viscous fingering instability, occurs at the interface of two fluids in a Hele-Shaw cell when a lower viscosity fluid is injected into a more viscous fluid, leading to a dynamic process of finger-like pattern formation at the fluid-fluid interface \cite{hill_channeling_1952,taylor_1958,paterson_1981,nye_interfaces_1984,chen_radial_1987}.  In contrast, when a higher (or equal) viscosity fluid invades a lower viscosity fluid, the interface is stable and forms a uniformly spreading front.  Many variations of this instability have been constructed such as those which destabilize conventionally stable Hele-Shaw flows through the implementation of variable geometry of the Hele-Shaw cell \cite{zhao_perturbing_1992, al-housseiny_control_2012,al-housseiny_controlling_2013}, the introduction of surfactants \cite{guo_surface-tension-driven_1992, guo_dynamics_1995, chan_observations_1997, chan_surfactant_2000, krechetnikov_new_2004, fernandez_experimental_2005,rocha_manipulation_2013}, or the presence of $A+B\to C$--type chemical reactions at the fluid interface \cite{almarcha_chemically_2010,mishra_influence_2010,riolfo_experimental_2012}.  One variation of interest is the introduction of a global rotation of the Hele-Shaw cell about a perpendicular axis through the center at a prescribed frequency \cite{carrillo_experiments_1996,Carvalho2014_HS_elasticA,alvarez-lacalle_nonlinear_2004,schwartz_instability_1989,alvarez-lacalle_low_2004}. If the density of the inner fluid exceeds that of the outer, the system is susceptible to a centrifugal instability resembling the well-known Rayleigh-Taylor instablity of superposed fluids as the denser fluid in the center is now unstable to outward displacement. In general the two fluids may have different viscosities such that the instability is mediated by competing centrifugal and viscous effects, or the viscosities may be taken to be the same so that only inertial effects drive the instability.  More recently, this instability has been considered in systems where the interface has additional properties such as a curvature-dependent bending modulus due to a chemical reaction between the two fluids or constrained length \cite{he_modeling_2012,Carvalho2013_HS_interfacial,Carvalho2014_HS_elasticA,Carvalho2014_HS_elasticB,foster_2021}. If the interface is taken to have nonzero bending modulus, the length scale of the emerging fingering patterns is set by the competition between the bending of the elastica and the centrifugal driving \cite{pocivavsek2008}.  The fingering structures which emerge can take the form of a periodic finger pattern commonly referred to as wrinkles, or isolated single or multiple localized structure(s) called folds. These states represent steady states of the system characterized by force and torque balance. Similar pattern formation occurs in many other systems beyond the Hele-Shaw geometry, for example, during the dynamic buckling of a membrane bounding a popped soap film \cite{box2020}, the dynamic wrinkling of a sheet due to drop impact \cite{box_dynamics_2019}, or the dynamic buckling of pressurized circular rings \cite{katifori_collapse_2009}.  Similar pattern formation may also occur quasistatically, for example, in the buckling of a ring due to a geometrically simple confinement \cite{hazel_buckling_2017}, compression or contact-induced wrinkling and folding of a floating elastic sheet \cite{roman_elasto-capillarity_2010,diamant_witten2011,king_elastic_2012,rivetti2013,LeoEdgarsheet}, three-dimensional deformations of water droplets under rotation \cite{brown_shape_1980}, the wrinkling and puckering of supported growing elastic struts \cite{michaels_geometric_2019,michaels_puckering_2021}, or the folding in biological systems such as airways \cite{heil_airway_2002,moulton_circumferential_2011}, ocular surfaces \cite{pournaras_macular_2000,neff_20_2013} or arteries \cite{pocivavsek2008,foster_2021}.


In this article, we examine the deformation of an elastic interface between two fluids in a rotating Hele-Shaw cell. This system has been studied in a series of papers by Carvalho et al \cite{Carvalho2013_HS_interfacial,Carvalho2014_HS_elasticB,Carvalho2014_HS_elasticA} on the assumption that the pressures in the interior and exterior fluids are identical and that the interface cannot support tension. The second assumption is consequential: it implies there is no force that resists changes in the interface length. As a result, Carvalho et al report a series of steady state profiles with different interface lengths without organizing these states into a bifurcation scenario that describes how the steady states of the system vary with system parameters.

We adopt here a different approach. We include an interface tension comprised of a curvature-dependent term and a Lagrange multiplier $T$ required by the assumed inextensibility of the interface, and refer to $T$, for simplicity, as the tension. This quantity is determined by solving a nonlinear eigenvalue problem and quantifies the response of the system to changes in the system parameters, at fixed interface length. It is therefore possible to plot the solutions in a $(P,T)$ plane, measuring the response of the system (the tension $T$) to changes in the pressure difference $P$ between the inner and outer fluids. If $P$ is held fixed and the rotation rate $\Omega$ is varied, one may instead show the $(\Omega,T)$ plane. Such diagrams are examples of bifurcation diagrams and they allow one to track changes in the solution profile as a parameter is varied since the profile is determined in the process of solving the eigenvalue problem for $T$. Numerical continuation techniques are ideally suited to this purpose and we use them here. In the case where $T\equiv0$, i.e. the length is unconstrained, we use the length $L$ of the interface as a new parameter and track how $L$ changes as a function of $P$ or, equivalently, of the rotation rate $\Omega$.

The steady states of the rotating two-fluid Hele-Shaw system are described by a model equation derived from force and torque balance using the simplest nonlinear bending energy possible for an interface with constant bending modulus. In previous work, we studied an equivalent equation describing the wrinkling of an elastic lining of an artery under compression and provided an explanation via weakly nonlinear analysis and numerical continuation for how primary and secondary solutions emerge and the forms they take \cite{foster_2021}. We also noted that the finger profiles of this system map onto the buckled states of an elastic ring under pure compression, albeit at different locations in parameter space \cite{foster_2022}. Since the latter problem is integrable in terms of elliptic functions, this is also the case for the present problem \cite{Arreaga_shapes,Djondjorov_equil_shapes,djondjorov_ICGIQ,foster_2022}.  It is not, however, the case for the secondary states comprising mixed modes and spatially localized folds whose properties we also investigate. The net result is a rather complete picture of the steady states of this interesting system.

This paper is organized as follows. In Sec.~\ref{system} we formulate the problem. Section \ref{constrained} introduces the length-constrained problem and summarizes the linear stability properties of a circular interface. Section \ref{unconstrained} summarizes parallel results for the case $T=0$. Section \ref{method} describes the formulation of our numerical continuation approach, and the numerical tests carried out to validate it. This is followed in Secs.~\ref{fingers}--\ref{folds} by the results tracking the properties of fingers, mixed modes and folds as a function of the parameters in the length-constrained case, followed by a summary of our results in the unconstrained case in Secs.~\ref{Tzero}--\ref{Tzerofolds}.  Finally, in Sec.~\ref{chiral} we consider asymmetric or chiral states and investigate their origin in parameter space. The paper concludes with a brief summary in Sec.~\ref{conclusion}.





\section{The system}\label{system}

\begin{figure}[t]
    \centering
    \includegraphics[width=0.5\linewidth]{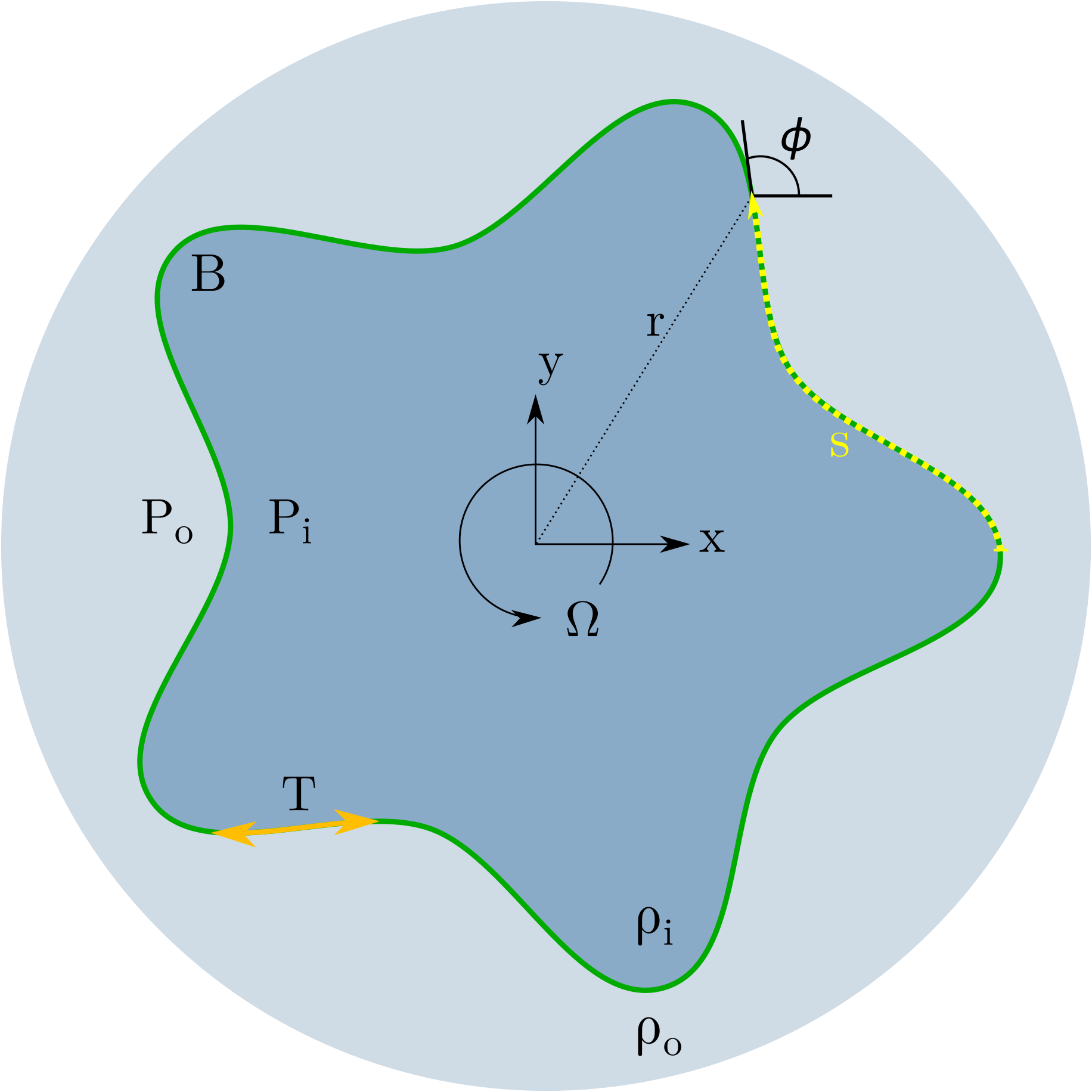}
    \caption{Top view of a rotating Hele-Shaw cell containing two fluids of densities $\rho_i$ and $\rho_o$ separated by an elastic membrane wrinkled with wave number $m=5$. The interface is parametrized by arclength $s$, with $(x(s),y(s))$ providing a parametric representation of the interface profile.}
    \label{fig:model1}
\end{figure}

We consider a Hele-Shaw cell with a fluid of density $\rho_i$ surrounded by a fluid of density $\rho_o$ and separated from it by a closed, elastic membrane of length $L=2\pi R$ and bending modulus $\mathcal{B}$, assumed to be independent of the curvature (Figure ~\ref{fig:model1}). The system rotates with constant angular velocity $\Omega$ about an axis perpendicular to the cell at $\textbf{r}=0$. In equilibrium, the inner liquid occupies a circular region with the axis of rotation at its center. We are interested in understanding the properties of this equilibrium state as either the rotation rate increases for a given pressure difference $P$ between the fluids, or as $P$ varies for a given rotation rate $\Omega$.

We parametrize the perturbed interface with the arclength $s$ such that the curvature of the interface at location $s$ or equivalently at $\textbf{r}(s) = (x(s),y(s))$ relative to the origin is given by $\kappa=\partial_s\phi$, where $\phi(s)$ is the angle between the tangent to the interface at point $s$ and the $x$ axis (Fig.~\ref{fig:model1}).  We define the density difference $\Delta\rho\equiv\rho_i-\rho_o$ between the interior and exterior fluids and impose a pressure difference $P\equiv P_i-P_o$ between them. A tension $T$ in the elastica is required to maintain its inextensibility. We call the resulting problem the \textit{constrained length} problem. In contrast, when $T$ is set to zero the length of the interface is unconstrained, and we call the resulting problem the \textit{unconstrained length} problem.  Both forms of the problem exhibit steady-state solutions, and it is solely these solutions we consider in this paper.

\section{Constrained Length}\label{constrained}

With the system thus defined, we can derive an equation for the interface by balancing the normal forces and the torque on an element of length $ds$ of the interface. The normal force includes contributions from the elastica bending modulus, tension and centrifugal force together with the force from the imposed pressure difference.  The resulting Kirchhoff equations for the interfacial elastica are given in the Supplementary Material~\cite{supp}, and can be manipulated to yield the governing equation
\begin{equation}
\mathcal{B}\left(\partial_{s}^{2}\kappa+\frac{1}{2}\kappa^{3}\right)-T\kappa-P-\frac{1}{2}\Delta\rho\,\Omega^2 r^{2}=0\,.\label{eq:main}
\end{equation}
Here $r(s)$ is defined implicitly via the geometrical constraints $\partial_{s}x=\cos\phi$ and $\partial_{s}y=\sin\phi$. In the unconstrained case the resulting equation is identical to that used in previous work on centrifugally driven instabilities in a Hele-Shaw cell \cite{Carrillo1996,Carvalho2014_HS_elasticA,Carvalho2014_HS_elasticB,alvarez-lacalle_nonlinear_2004}.

We expect a centrifugal instability to set in when $\Delta\rho>0$, i.e., when the density of the inner fluid exceeds that of the outer one, and use this fact to introduce the natural length scale of the instability
\begin{equation}
\lambda\equiv\left(\frac{\mathcal{B}}{\Delta\rho\,\Omega^2}\right)^{\frac{1}{5}}\,.\label{eq:natural_wavelength}
\end{equation}
We use this scale to construct a dimensionless parameter $\ell\equiv R/\lambda$ that measures the radius $R$ of the elastica in units of the natural length $\lambda$.

Rescaling Eq.~(\ref{eq:main}) according to $s\sim R$, $\kappa\sim R^{-1}$, $r\sim R$, $T\sim\mathcal{B}/R^{2}$, $P\sim\mathcal{B}/R^{3}$,
we obtain 
\begin{equation}
\partial_{s}^{3}\phi+\frac{1}{2}\left(\partial_{s}\phi\right)^{3}-T\partial_{s}\phi-P-\frac{1}{2}\ell^{5}r^{2}=0\,.\label{eq:phiODE}
\end{equation}
Instability of the circular membrane arises when the denser interior fluid is displaced outwards, thereby increasing the outward force upon it. When this force exceeds the restraint arising from the curvature of the interface, the tension $T$, and the imposed pressure difference $P$, instability sets in and it is this balance which gives rise to the wavelength selection.

\begin{figure}[h!]
\centering{}\includegraphics[width=0.75\linewidth]{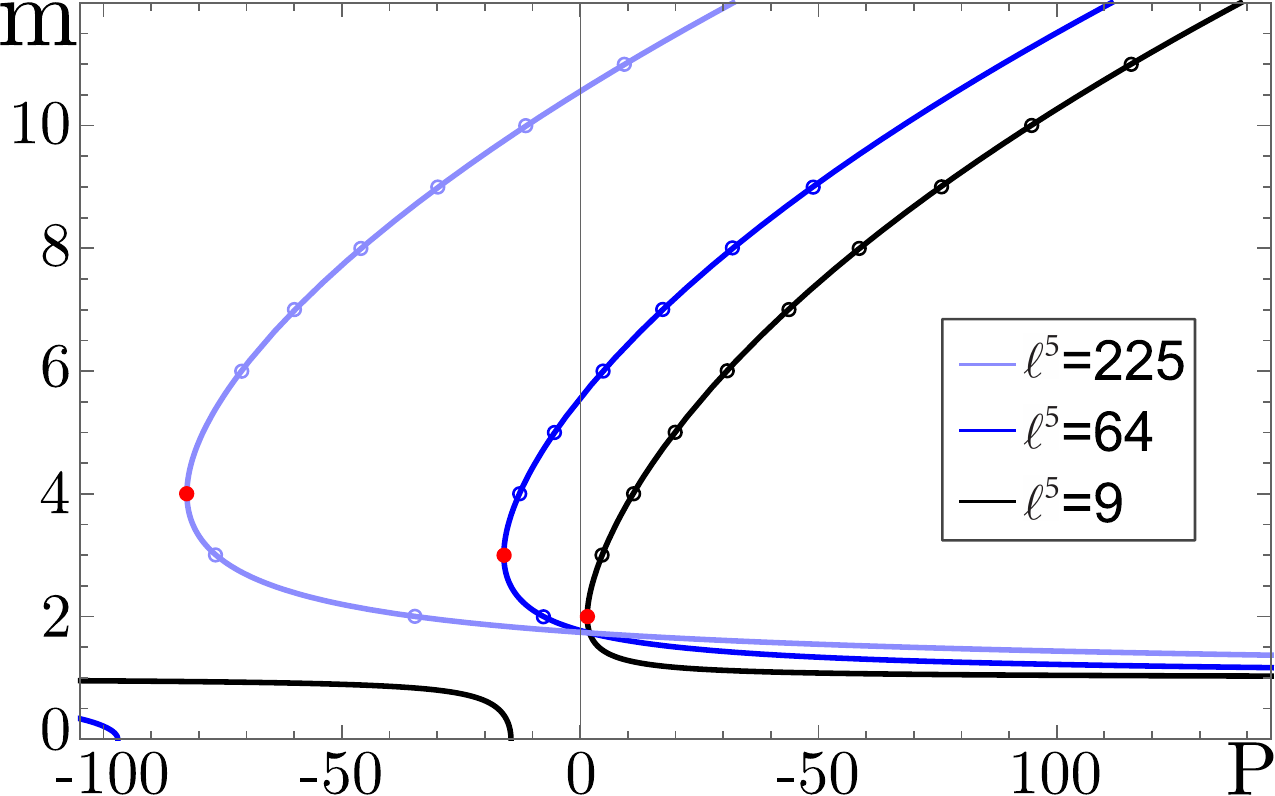}
\caption{The perturbation wave number $m$ as a function of $P$ across a range of $\ell^5$ values corresponding to onset wave numbers $m^*=2,3,4$ (red dots). As $\ell^5$ increases, instability sets in at lower values of $P$ and with larger values of $m^*$, i.e., larger finger wave numbers. Since $L=2\pi$ the $\ell^5$ values have been chosen to yield integer wave numbers at onset.\label{fig:P0}}
\end{figure}

In the unperturbed problem, the interface is circular and of length $L=2\pi$, with $T=\frac{1}{2}(1-\ell^5)-P$. A linear stability analysis of this state yields the dispersion relation for the mode number $m$ of the fingering instability \cite{foster_2021}
\begin{equation}
    m^4-\left(2+P+\frac{\ell^5}{2}\right)m^2+\left(1+P+\frac{3}{2}\ell^5\right)=0.
    \label{disp1}
\end{equation}
From this relation, we can determine the wave number $m^*$ of the first unstable mode that sets in as $P$ increases and the critical value $P=P^*$ at which it does so: 
\begin{align}
  m^{*} & =(1+\ell^{5/2})^{1/2}\,, \qquad P^{*}\equiv(-\ell^{5}+4\ell^{5/2})/2\,,\label{eq:mc1}
\end{align}
all for a fixed rotation rate $\Omega$. Alternatively, we may fix $P$ and increase the rotation rate leading to the critical rotation rate $\Omega(m,P)$,  
\begin{align}
    \Omega & \equiv \left[\frac{B}{\Delta \rho R^5} \left(\frac{2(m^2-1)(P-m^2+1)}{3-m^2} \right) \right]^{1/2}\,, \label{eq:mc2}
\end{align}
for the appearance of a mode with wave number $m$; minimizing this expression over $m$ for fixed $P$ recovers the results in Eq.~(\ref{eq:mc1}). Figure~\ref{fig:P0} shows several examples of the marginally stable wave number $m$ as $P$ and $\ell^5$ vary and shows that, as $\ell^5$ increases, instability sets in at lower and lower values of $P$ and with larger and larger values of $m^*$, i.e., larger finger wave numbers. This is the fingering instability whose nonlinear development is key to understanding the constrained system. When $m^{*}$ is not an integer, the figure shows that the primary instability corresponds to the integer $m$ nearest to the $m^{*}$ given by Eq.~(\ref{eq:mc1}).  

\section{Unconstrained Length}\label{unconstrained}

When the interface cannot support any tension, the instability may lead to interface growth. In this case $T=0$ and the length $L$ of the interface becomes a free parameter. We are interested in this case because earlier work by Carvalho et al.~\cite{Carvalho2014_HS_elasticB} presented a number of solutions to Eq.~(\ref{eq:phiODE}) with $T=0$ with interfaces of different (and unspecified) lengths. With $L$ as a free parameter, nondimensionalization requires rescaling Eq.~(\ref{eq:main}) using the natural length scale $\lambda$.  We take $s\sim \lambda, \kappa\sim \lambda^{-1},r \sim \lambda, P \sim \mathcal{B}/\lambda^3$ such that the rescaled governing equation is now given by 
\begin{equation}
    \partial_s^2 \kappa + \frac{1}{2}\kappa^3 - P - \frac{1}{2} r^2 = 0.
    \label{eq:T0}
\end{equation}
In the case of a circular solution, this yields the (monotonic) relation between the pressure and interface length: $P = ((L/2\pi)^{-3}-(L/2\pi)^2)/2$, so in equilibrium there is only one $L$ for a given $P$.  

A linear stability analysis of this state, similar to that leading to (\ref{disp1}), yields
\begin{equation}
  m^4-\frac{5}{2} m^2 + \left[\frac{3}{2} + \left(\frac{L}{2\pi} \right)^4\right] = 0\,,
  \label{disp2}
\end{equation}
after dividing out the marginally stable mode $m=0$. This relation has no real roots for $L\ge \pi$, indicating that solutions of wave number $m$ do not emerge as primary bifurcations from the circle state as $P$ varies.

\section{Method}\label{method}

Both versions of the problem, Eqs.~(\ref{eq:phiODE}) and (\ref{eq:T0}), are implemented in AUTO \cite{doedel08auto-07p} as a 5-dimensional boundary value problem encompassing the third order ODE for $\phi$ and the two first order ODEs for $(x,y)$.  For the constrained length sections of this paper, we construct the problem on the domain $s\in[0,\pi]$, representing half of a closed elastic interface of length $2\pi$, subject to the boundary conditions
\begin{subequations}
\begin{align}
& \phi(0) =\pi/2\,, \quad \phi(\pi)=3\pi/2\,,\\
& x(0)=x_{0}\,, \quad x(\pi)=x_{1}\,, \\ & y(0)=y(\pi)=0\,, \\
& \partial_s^2\phi(0)=\partial_s^2\phi(\pi)=0\,.
\end{align}
\label{bcs}
\end{subequations}
The final two conditions represent force-free conditions following previous work \cite{kodio_weak_nonlin,foster_2021} and allow us to generate the full circle solution via reflection in the $x$ axis. For the unconstrained length sections, $s\in[0,L/2]$ where $L$ is free to vary.  Since the system is 5-dimensional with 8 boundary conditions, numerical continuation is performed in the four parameters $(P,T,x_{0},x_{1})$ for the constrained length problem and $(P,L,x_{0},x_{1})$ for the unconstrained length problem \cite{Doedel_1991}, i.e., for a given change in $P$, the new tension $T$ is found as a nonlinear eigenvalue of the problem, while $x_0,x_1$ are adjusted to satisfy the force-free boundary conditions.

The results of this procedure have been carefully compared to those from a high order weakly nonlinear analysis and are found to be in excellent agreement \cite{foster_2021}. The nonlinear results have also been confirmed to match exact analytical solutions to within numerical tolerance \cite{foster_2022}.  Self-contact forces can be included in order to continue solutions beyond our current range \cite{flaherty_1972,flaherty_1973}, but we do not do so in this paper. Self-intersecting solutions are not shown.

The boundary conditions (\ref{bcs}) impose a reflection symmetry on all the solutions generated by the above procedure and so prevent the computation of states that break this symmetry. Since such states are also expected to be present \cite{Carvalho2014_HS_elasticB}, we discuss in the penultimate section, Sec.~\ref{chiral}, an alternative formulation of the above problem that permits the computation of such solutions.





\section{Constrained length: fingers}\label{fingers}

\begin{figure}[h!]
\centering{}\includegraphics[width=0.99\linewidth]{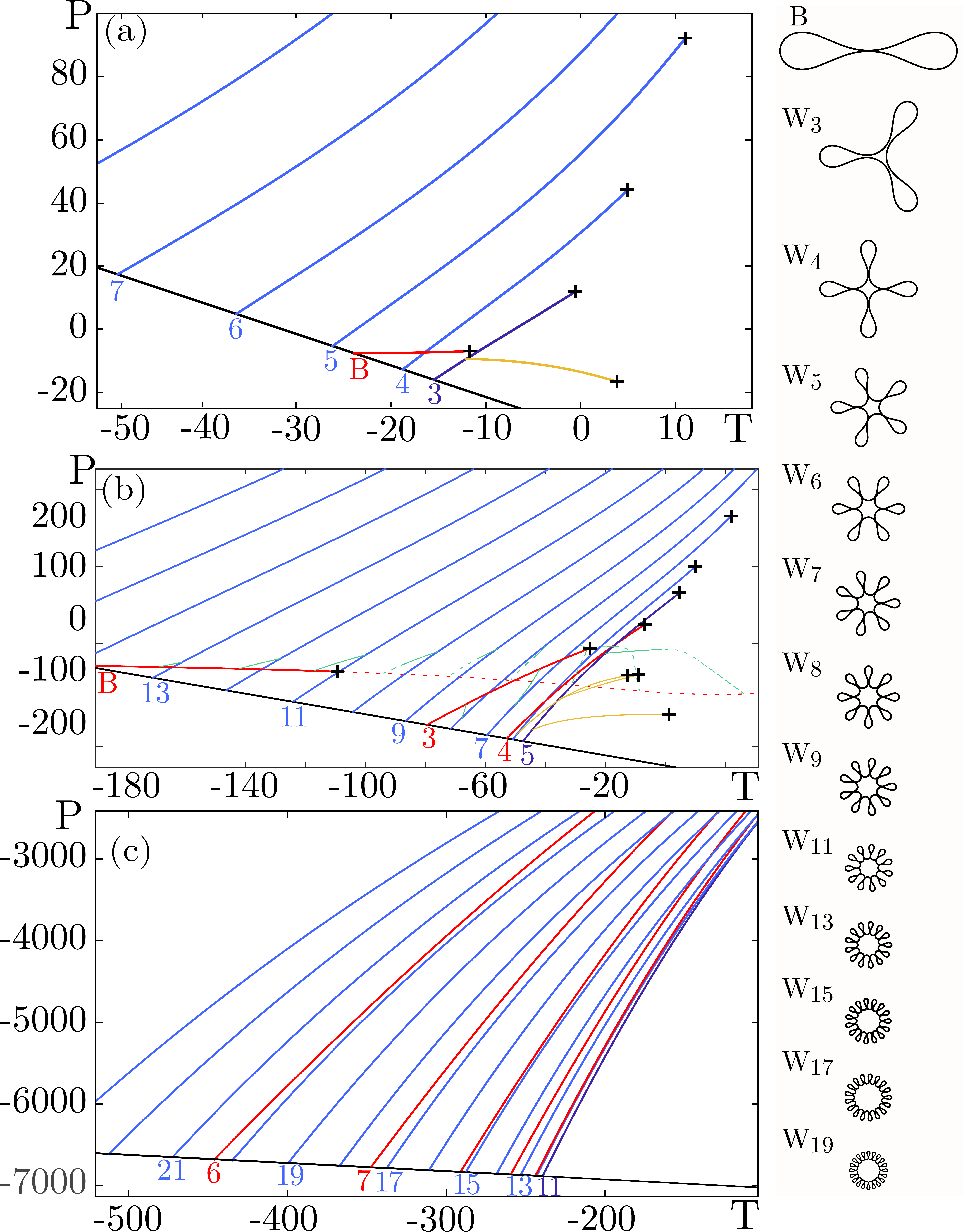}
\caption{Bifurcation diagrams in the $(T,P)$ plane for (a) $\ell^5=64$ ($m^*=3$), (b) $\ell^5=576$ ($m^*=5$) with some states beyond self-contact shown with dashed lines, and (c) $\ell^5=14400$ ($m^*=11$), showing the succession of color-coded primary branches of finger states as $P$ increases: primary fingers with $m=m^*$ (purple), $m>m^*$ (blue), and $m<m^*$ (red). Mixed-mode states are shown in (a) and (b) (green), together with branches of folds (yellow).  In (c) only the primary finger modes are shown.  Primary bifurcation points are labeled by the corresponding wave number, while points of self-contact are indicated with crosses (for $m\le 7$ in panel (b)); in panel (c) points of self-contact fall outside the parameter range shown.  Selected solution profiles at the point of self-contact are shown on the right.
\label{fig:BD35911}}
\end{figure}

The order with which primary finger states bifurcate from the circle state is determined by the parameter $\ell^5$ as it sets the critical wave number $m^*$ via Eq.~(\ref{eq:mc1}).  Subsequent bifurcations as $P$ increases lead to modes with wave numbers $m$ alternately above and below $m^*$. Once $m=2$ is reached only modes with large wave numbers $m$ remain and these continue to be destabilized as $P$ increases as seen in Fig.~\ref{fig:BD35911}. The figure shows the response of the system, as indicated by the tension $T$, to changes in the imposed pressure difference $P$ for three values of $\ell^5$ obtained using numerical continuation starting from the neutral modes of the circle state (black line). The corresponding solution profiles at the point of first self-contact are shown alongside. We see that for large enough $\ell^5$ the primary finger states bifurcate to secondary branches of mixed-mode states connecting a primary branch with $m<m^*$ to a primary branch with $m>m^*$ (Fig.~\ref{fig:BD35911}(b), shown in green), as well as to circumferentially localized states we call folds (Fig.~\ref{fig:BD35911}(a,b), shown in yellow) which do not connect to another branch. The mixed modes do not set in prior to self-contact when $\ell^5$ is small but begin to proliferate with increasing $\ell^5$; for this reason they are omitted from panel (c). The primary mode with $m=2$ is called here a buckling mode (labeled $B$) because of its radically different behavior at large $\ell^5$, with negative modulus $dP/dT$, behavior typical of buckling processes \cite{foster_2021}.  In the absence of a natural wavelength $\lambda$ or as $\lambda$ becomes large relative to the domain, the $m=2$ mode is the first to become unstable \cite{flaherty_1972}.

Single folds bifurcate from the first primary branch ($m=m^*$) and do so prior to self-contact whenever $m^*\ge 3$. Figure~\ref{fig:BD35911}(a) shows the case $m^*=3$ ($\ell^5=64$) and shows that these states emerge in pairs, here a single protrusion fold $F_{s^+}$ and a single intrusion fold $F_{s^-}$ (for profiles see Figs.~\ref{fig:LS2} and \ref{fig:LS}). The resulting branches track closely but do not self-contact at the same point; mixed-mode branches are absent.  In panel (b), for $\ell^5=576$, fingering, buckling and mixed modes are all present, as well as additional folds.  In (c), for $\ell^5=14400$, the number of mixed modes and folds becomes large, and these states are omitted.

The finger profiles are $\ell^5$-independent \cite{vassilev_cylindrical_2008,Djondjorov_equil_shapes,djondjorov_ICGIQ,foster_2022}, so the displayed self-contact solutions hold for any of the bifurcation diagrams, although their location in the $(T,P)$ plane does vary with $\ell^5$ \cite{foster_2022}.

\FloatBarrier

\subsection{Constrained length: $P=0$ fingers}
\begin{figure}[h!]
    \centering
    \includegraphics[width=0.95\linewidth]{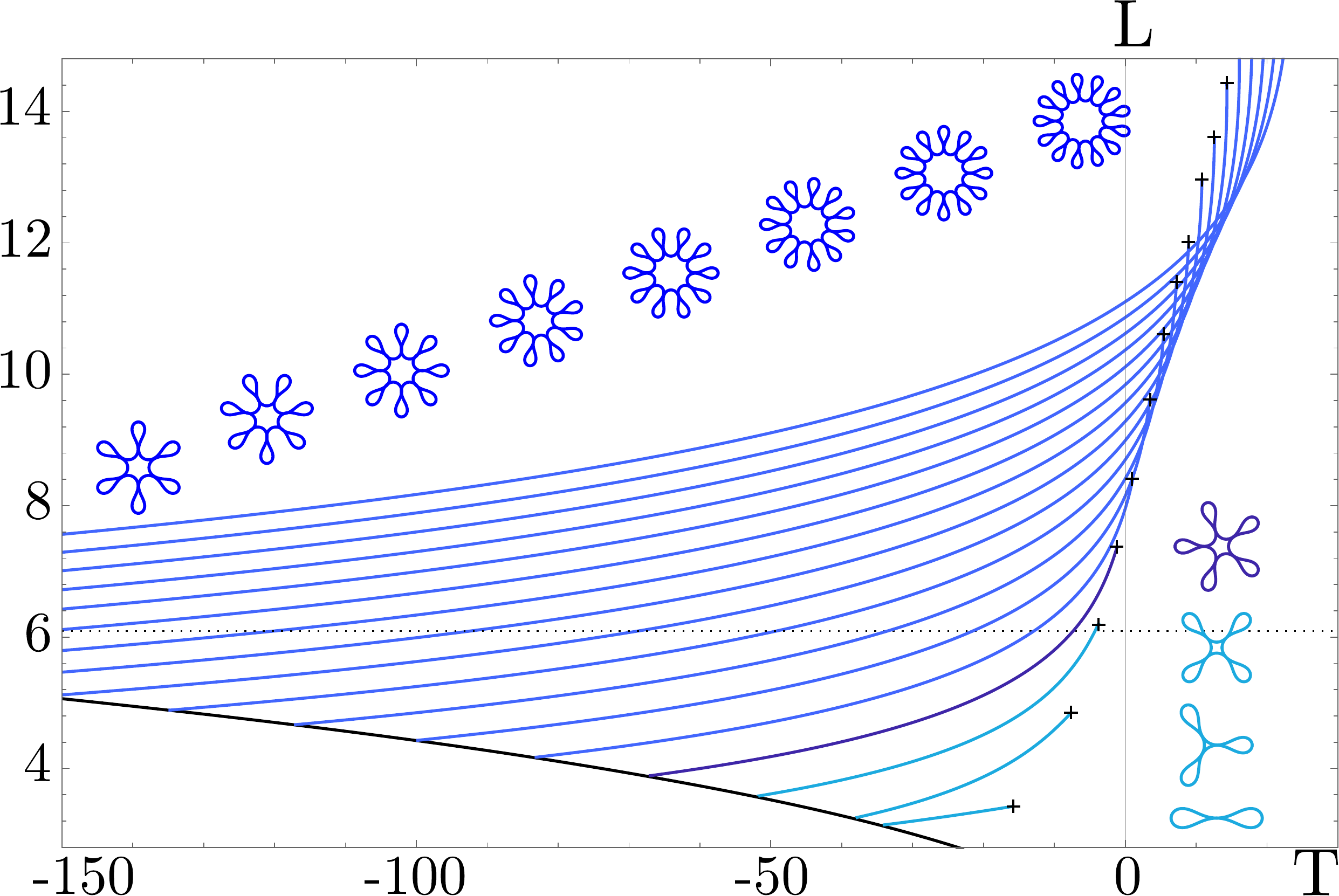}
    \caption{The $P=0$ steady state solutions in the $(T,L)$ plane for $\ell^5=576$ ($m^*=5$ in purple), where $L$ is the interface perimeter.  All profiles are plotted on the same scale. The horizontal dashed line indicates $L=2\pi$, the length used in all other constrained problem computations.}
    \label{fig:TL2}
\end{figure}

Finger solutions when the applied pressure $P=0$ are of physical interest as they dictate solutions which may be observed in ambient conditions, emerging solely from the competition between inertial and elastic energies.  Of course, these solutions have the same shape as before, but the length and tension at which they arise are not known {\it a priori}. Figure \ref{fig:TL2} shows their location in the $(T,L)$ plane.  Solutions bifurcate with $m$ increasing monotonically from $m^*=2$ as $L$ increases.  We note that the size of the marginally stable circle state may be smaller than in Fig.~\ref{fig:BD35911} for finger solutions with $m$ values close to $m^*$ since these typically bifurcate from the circle state already at more negative $P$, i.e. for these states $L<2\pi$. Moreover, for the value of $\ell^5$ used in Fig.~\ref{fig:TL2}, $\ell^5=576$, the states with $m=2,3$ reach self-contact at $L<2\pi$ and at negative values of $T$, i.e., under compression.
In contrast, for larger values of $m$, the circle at threshold is generally larger than $L=2\pi$ and self-contact generally occurs at small or slightly positive values of the tension $T$.


\FloatBarrier

\section{Constrained length: Mixed modes}

\begin{figure}
\centering{}\includegraphics[width=0.95\linewidth]{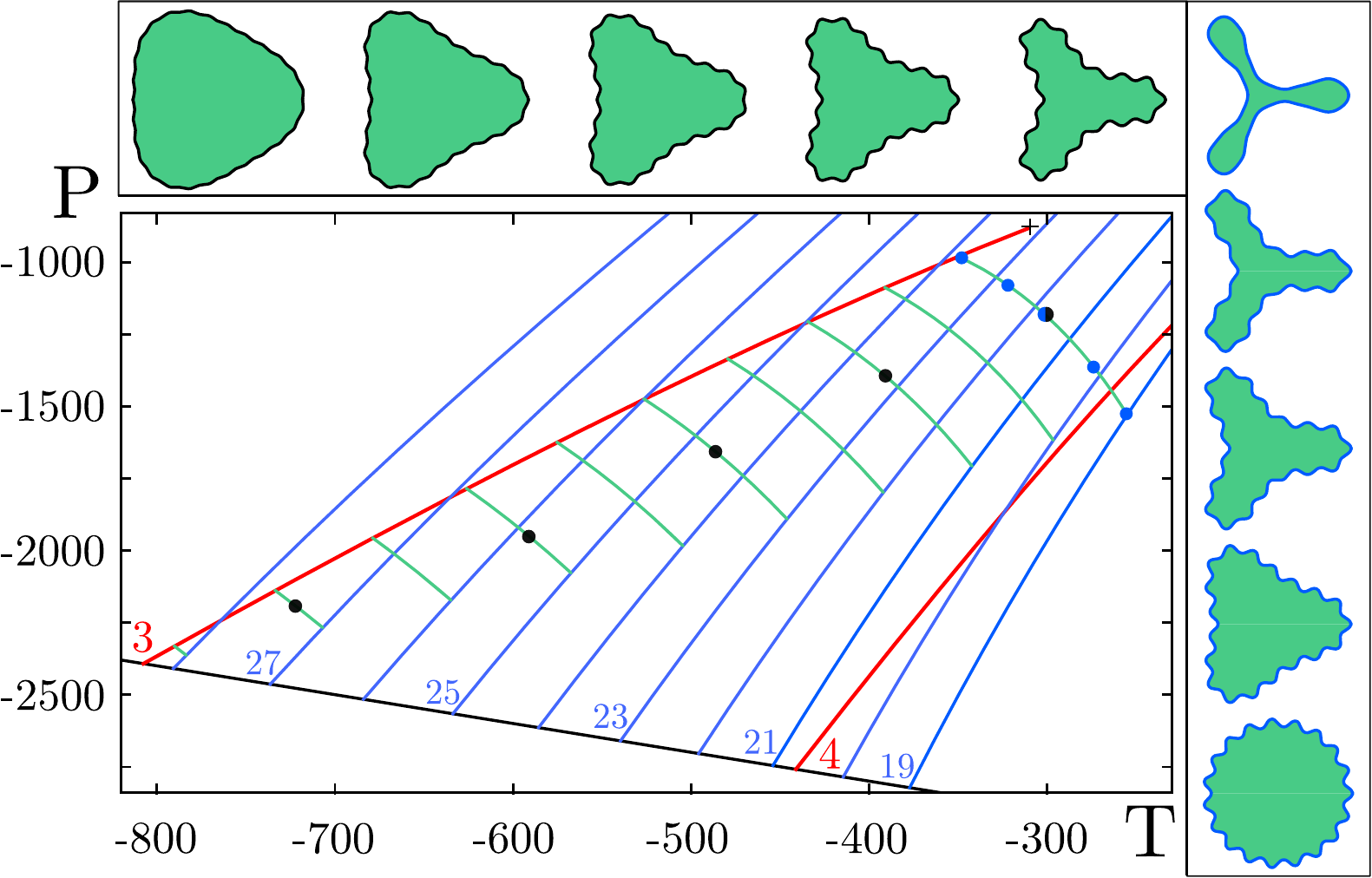}
\caption{A partial bifurcation diagram for $\ell^5=6400$ ($m^*=9$) with all complete mixed-mode connections between the $m=3$ finger state and other primary branches.  The profiles above the bifurcation diagram correspond to the black points proceeding from left to right, while the profiles on the right correspond to the blue points on the mixed-mode branch M$_{3,19}$, proceeding downwards. \label{fig:MM9}}
\end{figure}

In Figure~\ref{fig:MM9}, we show an example of the richness of the mixed-mode connections at higher values of $\ell^5$, here $\ell^5=6400$ ($m^*=9$), where more states with $m<m^*$ become closely interspersed with states with $m>m^*$.  The profiles above the diagram provide examples of mixed-mode states with comparable contributions from both wave numbers (black dots).  In contrast, the profiles along the side correspond to profiles along a single mixed-mode branch, here the branch M$_{3,19}$ connecting the $m=3$ and $m=19$ primary finger branches.


\begin{figure}
\centering{}\includegraphics[width=0.9\linewidth]{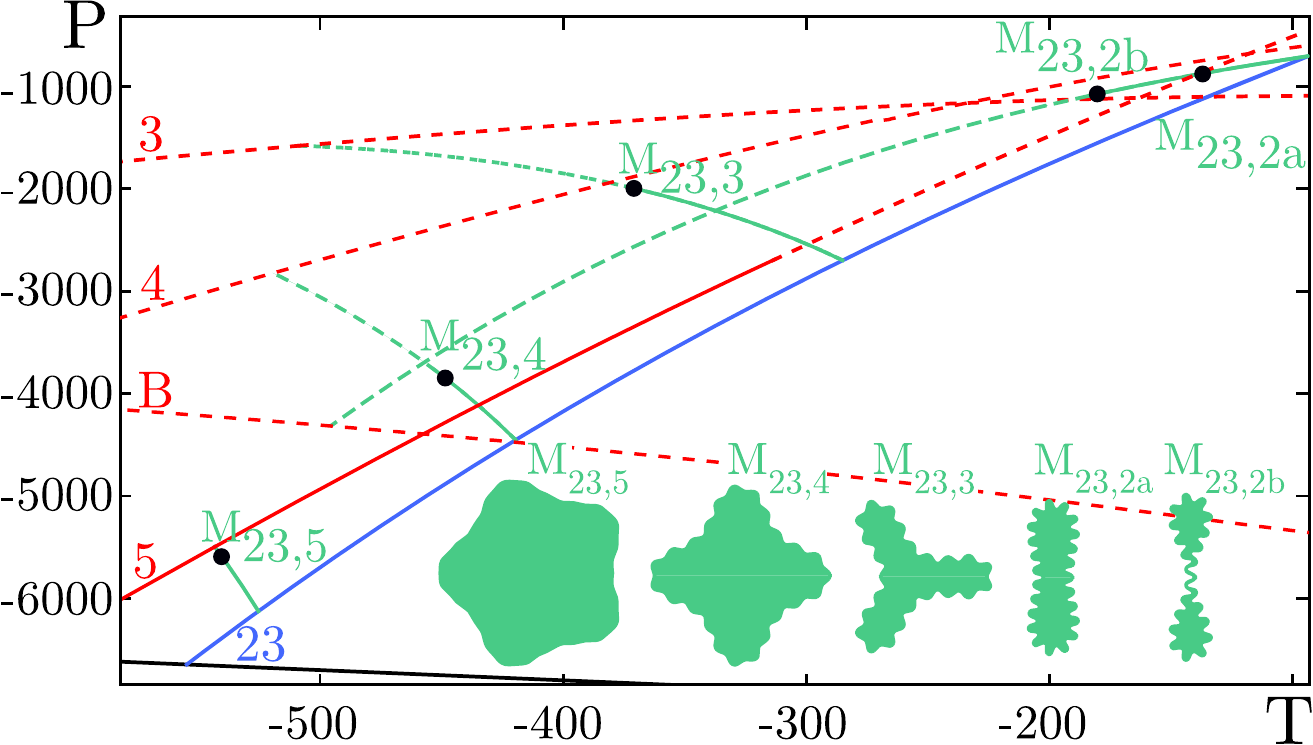}
\caption{A partial bifurcation diagram for $\ell^5=14400$ ($m^*=11$) with four mixed-mode connections between $m=23$ and the primary branches $m=2 \text{ (B)}, 3, 4, 5$.  Solutions beyond the point of self-contact are shown with broken lines. The lower panels show profiles at the locations indicated by black dots (two on the M$_{23,2}$ branch). \label{fig:MM11}}
\end{figure}

In Fig.~\ref{fig:MM11}, we show another example of mixed-mode connections. The figure shows the connections along a primary branch with $m=23$, i.e., $m>m^*$, as opposed to Figure~\ref{fig:MM9} where we showed all the secondary connections for a $m=3$ primary branch, $m<m^*$.  The solutions in Figs.~\ref{fig:MM9} and~\ref{fig:MM11} match closely those previously identified via a different computational procedure (see Figure 4 of Ref.~\cite{Carvalho2014_HS_elasticB}).  

\newpage

\section{Constrained length: Folds}\label{folds}

Additional secondary branches which do not connect to any other branches bifurcate from the finger branches at or near the critical branch with $m=m^*$ (Fig.~\ref{fig:BD35911}(a,b), yellow curves).  These correspond to localized solutions and typically come in pairs. The first pair always has a single intrusion or protrusion ($\mathrm{F}{}_{s^{\pm}}$) and bifurcates from the critical finger branch (Fig.~\ref{fig:BD35911}, yellow curves). Intruding and protruding states follow the same path in the bifurcation diagram and reach self-contact at almost the same point in the $(T,P)$ plane (Fig.~\ref{fig:BD35911}(a)).

\begin{figure}
\centering{}\includegraphics[width=0.95\linewidth]{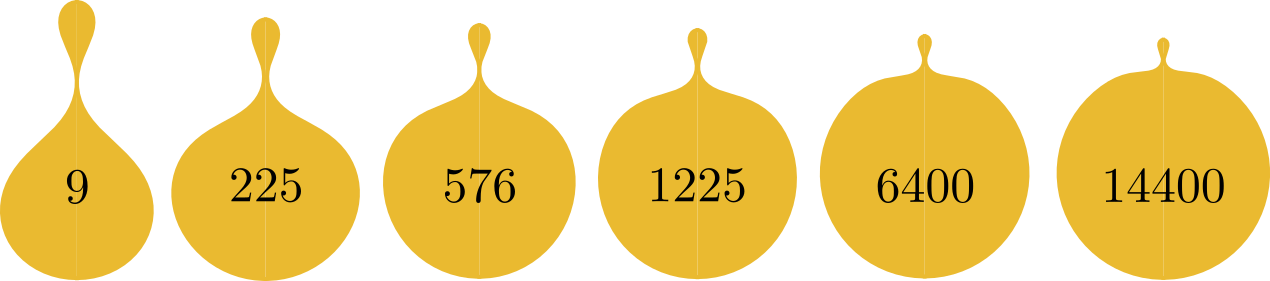}
\caption{$F_{s^+}$ solutions at self-contact for increasing values of $\ell^5$ corresponding to $m^*=3,4,5,6,9,11$. \label{fig:LS2}}
\end{figure}

Figure \ref{fig:LS2} explores the $\ell^5$-dependence of the simplest fold states F$_{s^+}$ at self-contact and shows that the width of the localized structure decreases in the expected way relative to the perimeter length as $\ell\equiv R/\lambda$ or equivalently the rotation rate increases.

\begin{figure}
\centering{}\includegraphics[width=0.95\linewidth]{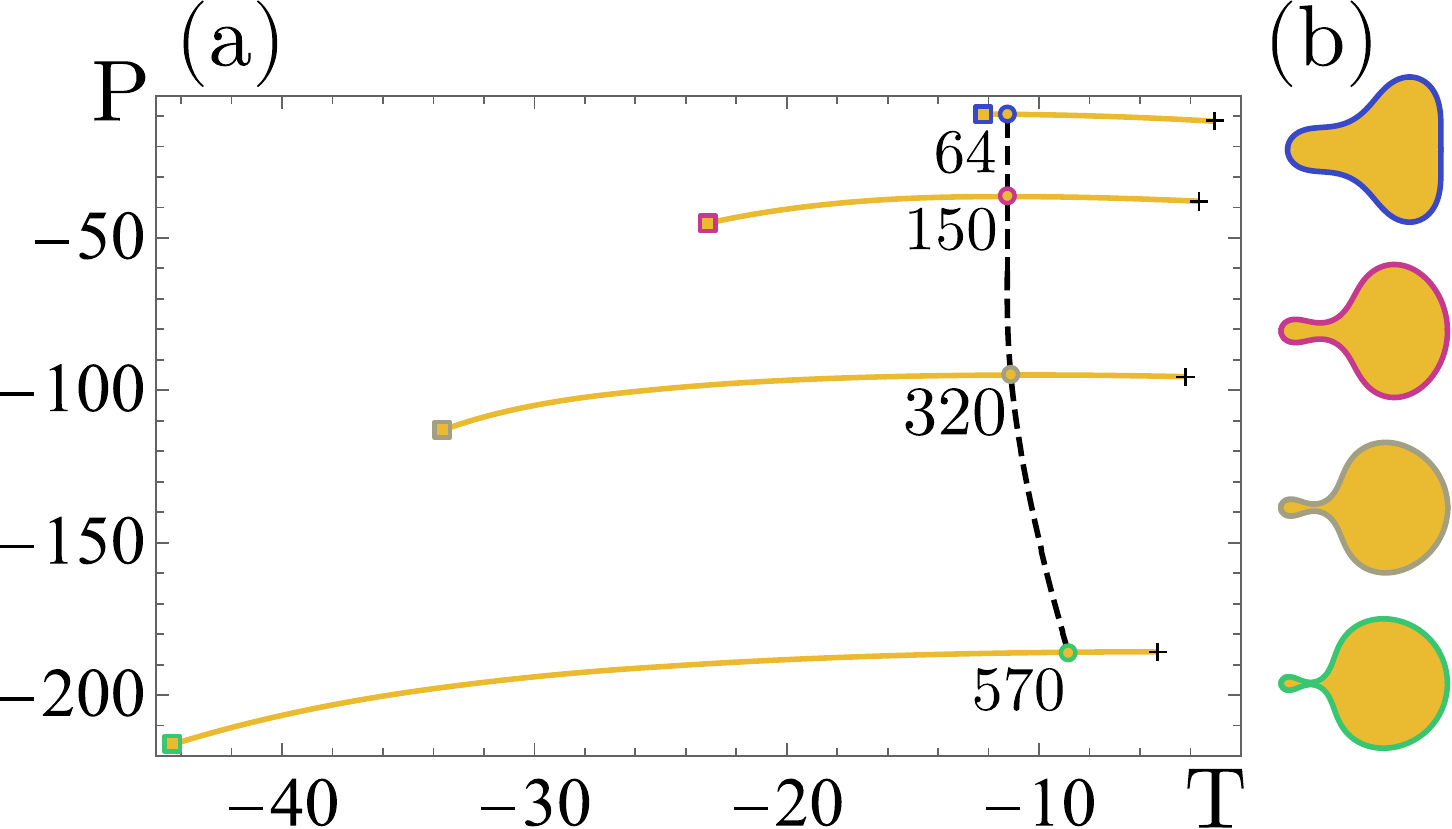}
\caption{Numerical continuation in the $(T,P)$ plane of the maxima along the branch of F$_{s^+}$ localized fold states starting at $\ell^5=64$ ($m^*=3$) as $\ell^5$ increases as indicated in the labels. Since $T$ measures the response of the system to changes in the pressure difference $P$, these maxima represent saddle-node bifurcations. The continuation is terminated when the saddle-node reaches the point of self-contact (crosses) or the bifurcation point from the primary branch (squares). The corresponding solution profiles at the labeled saddle-nodes are shown alongside.
\label{fig:LS}}
\end{figure}

Changes in $\ell^5$ also have an effect on the localized state branches in parameter space.  Figure~\ref{fig:LS}(a) shows that for moderate values of $\ell^5$ the branches of $\mathrm{F}{}_{s^{\pm}}$ fold on themselves, each exhibiting a saddle-node bifurcation. These bifurcation points can be followed numerically and Fig.~\ref{fig:LS}(b) shows the result of such a computation for $\mathrm{F}{}_{s^+}$ starting from the saddle-node bifurcation at $\ell^5=225$ and both increasing and decreasing $\ell^5$. Evidently, as $\ell^5$ varies, so does the wave number $m^*$ of the branch from which the F$_{s^+}$ bifurcate. This is in fact a continuous process: as $\ell^5$ increases, for example, the secondary bifurcation to F$_{s^+}$ moves down along the $m^*=4$ branch to the primary bifurcation point. As this happens, the $m=5$ primary bifurcation passes through the $m^*=4$ primary bifurcation point, so that for larger $\ell^5$ the circle state loses stability first to $m^*=5$, followed by $m=4$, i.e., the $m=4$ and $m=5$ branches exchange positions. Beyond this point the secondary bifurcation to F$_{s^+}$ moves up the new $m^*$ branch, and the whole process repeats \cite{LeoEdgarsheet,dangelmayr}. However, despite the jumps in $m^*$ the movement of the saddle-node of the fold states F$_{s^+}$ as $\ell^5$ varies is continuous.  With increasing $\ell^5$, the saddle-node moves past the point of self-contact, and we terminate the continuation when this first happens ($\ell^5 \approx 570$).


\begin{widetext}
\onecolumngrid\
\begin{center}\
    \begin{figure}[b!]
    \centering\includegraphics[width=0.85\linewidth]{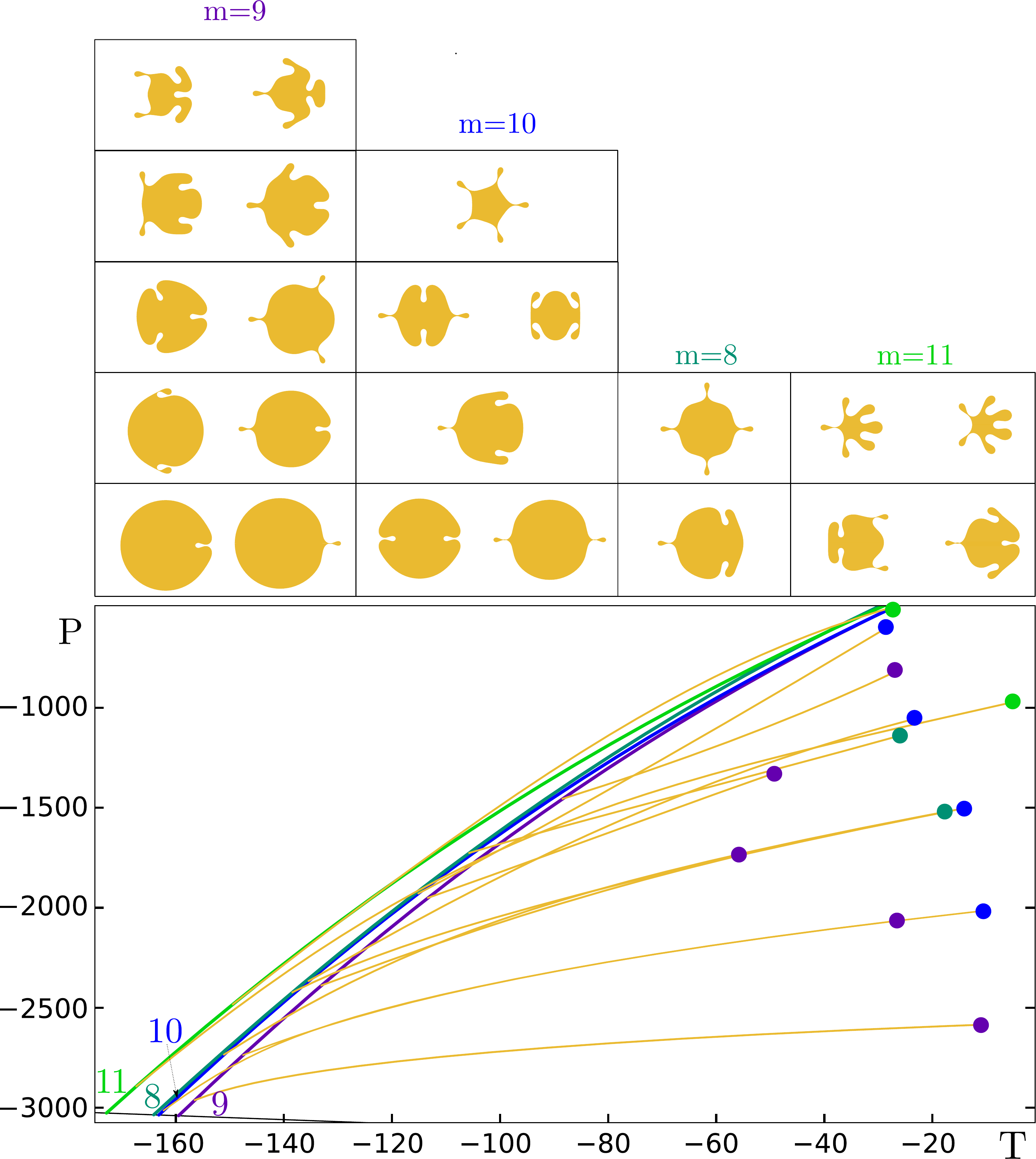}
    \centering\caption[width=0.5\linewidth]{Bifurcation diagram for $\ell^5=6400$ depicting the first few finger modes ($m^*=9$, and $m=10,8,11$, color-coded) and the multitude of remarkable fold states that bifurcate from them (yellow). Solutions at the point of self-contact, arranged in ascending order and labeled with dots of the corresponding color, are shown at the top. Branches of intrusions and protrusions are almost degenerate leading to overlapping dots. \label{fig:m9LS}}
    \end{figure}
\end{center}\
\twocolumngrid\
\end{widetext}

As $\ell^5$ increases, the variety of localized fold states increases dramatically since fold states now bifurcate not only from the critical finger state with $m=m^*$, but also from the subsequent primary finger branches.  In Fig.~\ref{fig:m9LS}, we show all localized fold states for a given $\ell^5$  ($\ell^5=6400$).  The critical primary branch supports the most secondary bifurcations to fold solutions. Although subsequent branches have a similar overall number of secondary bifurcations, the number of bifurcations to fold states decreases while the number of bifurcations to mixed-mode states increases. The details depend on the number of factors of each integer $m$ and hence the symmetry of the branch. For example, Fig.~\ref{fig:m9LS} shows several localized states with a $\frac{2\pi}{3}$ rotational symmetry emerging from the $m=9$ finger branch which shares this symmetry.  Similarly, we see a $\frac{2\pi}{5}$ rotational symmetry in one of the localized states emerging from $m=10$ and a $\frac{\pi}{2}$ rotational symmetry in one that emerges from $m=8$.


\section{Unconstrained length: Fingers}\label{Tzero}

\begin{figure}[h!]
    \centering
    \includegraphics[width=0.9\linewidth]{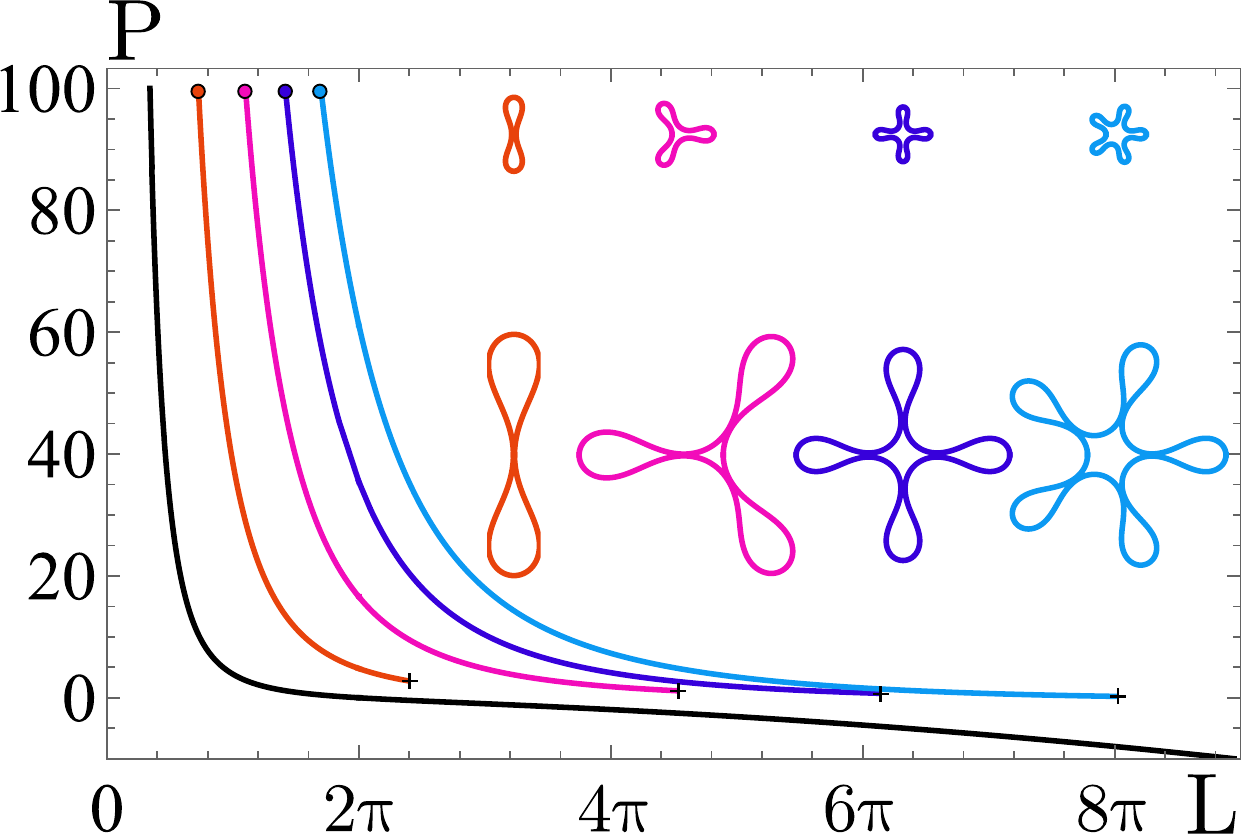}
    \caption{Numerical continuation of $m=2,3,4,5$ finger states (orange, pink, dark and light blue, respectively) in the $(L,P)$ plane. The branch of circle states is shown in black, and self-contact is indicated by crosses. The finger states no longer bifurcate from the circle state. \label{fig:PL1}}
\end{figure}

We now consider the same system but this time take $T$ to be zero and the interface length $L$ to be a free parameter, resulting in Eq.~(\ref{eq:T0}). Numerical continuation for steady-state symmetric finger states may be performed in the same way as before. Figure~\ref{fig:PL1} shows the result of continuing several finger states in the $(L,P)$ plane.  We observe that as the interior pressure drops, the area and perimeter of the steady-state finger profile grow dramatically.  With a smaller contribution from internal pressure, a larger contribution from the centrifugal force is needed to support a steady state, requiring larger $L$.  In contrast to the results with nonzero tension, in this case the finger states no longer bifurcate from the circle state. Moreover, although secondary bifurcations do set in as $L$ increases, for these small values of $m$ they do so far beyond the point of first self-contact. Figure~\ref{fig:MM_T0} shows that this is no longer so for larger values of $m$.

\section{Unconstrained length: Mixed modes}\label{Tzerommodes}

\begin{figure}
    \centering
    \includegraphics[width=0.99\linewidth]{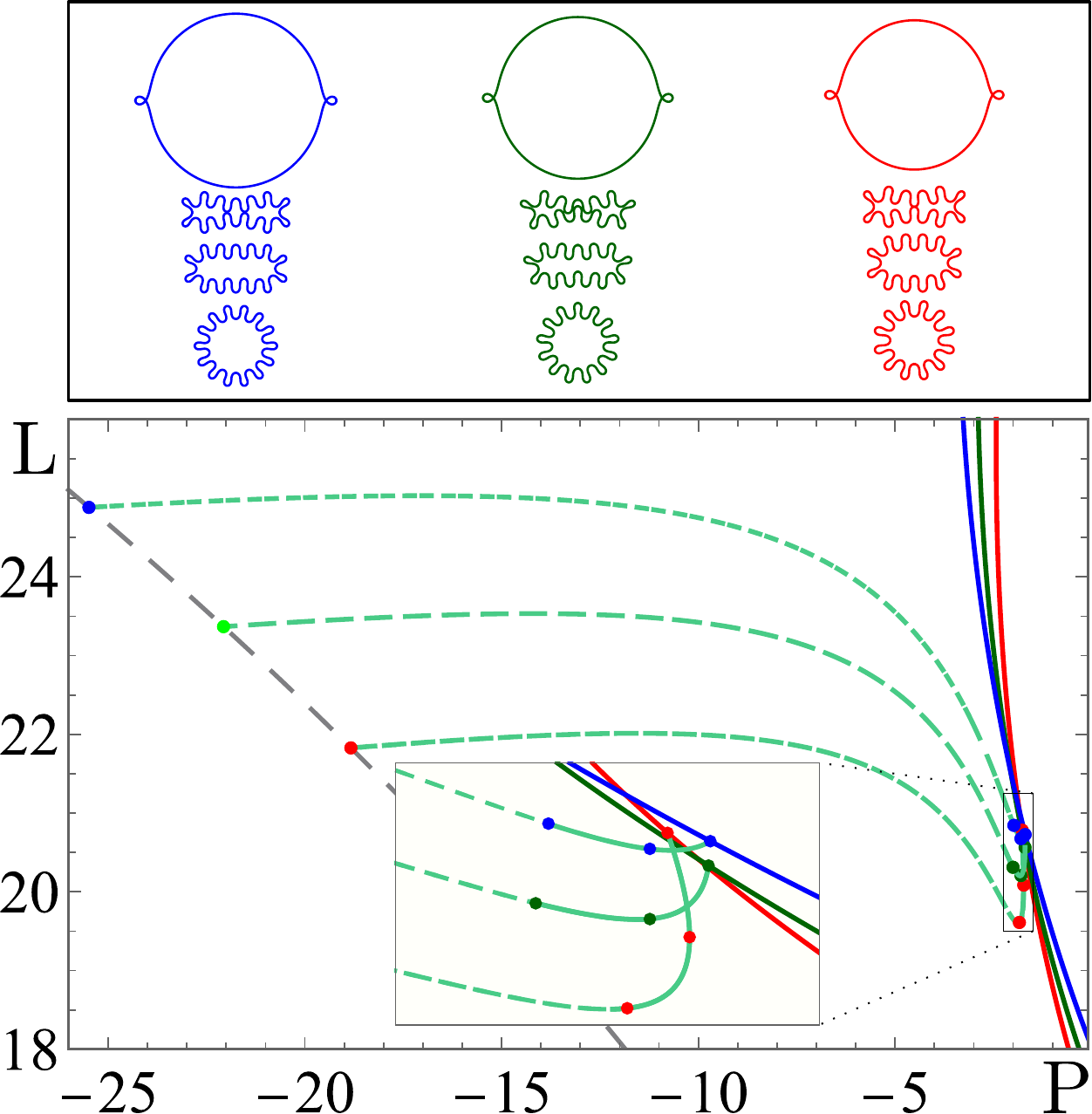}
    \caption{Bottom: mixed-mode connections (light green) between $m=2$ (grey) and $m=12$ (red), $m=13$ (dark green) and $m=14$ (blue) finger states shown in the $(P,L)$ plane. Dashed lines indicate solutions beyond first self-contact. Top: sample solutions at the color-coded locations indicated in the bottom panel proceeding from the $m=2$ start of the mixed-mode branch to its endpoint at the other end. In each case the top profile corresponds to an unphysical, self-intersecting solution on the $m=2$ branch while the profile below corresponds to last self-contact (lower panel zoom). Beyond this point the solutions are realizable; the final profile corresponds to a pure finger state at the right end of the branch. \label{fig:MM_T0}}
\end{figure}

Mixed-mode states connecting two primary branches with different numbers of fingers can also be found. Figure~\ref{fig:MM_T0} shows three mixed-mode branches (light green lines) originating on the $m=2$ finger branch (grey line), shown in the $(P,L)$ plane. The branches terminate on the $m=12$ (red), $m=13$ (dark green) and $m=14$ (blue) finger states, respectively. In each case the solutions are realizable near the high $m$ end of the branch (solid light green lines) but as $P$ becomes more and more negative they make self-contact beyond which the solutions are no longer realizable. The top panel in the figure shows the pure finger state at the right end of each branch (lowest profile), a physical solution before self-contact (second profile from the bottom), the solution at first self-contact (third profile from bottom) and finally an unphysical, self-intersecting profile very close to the termination of the mixed-mode branch on the $m=2$ branch (top profile). 

\section{Unconstrained length: Folds}\label{Tzerofolds}

Fold states arise through bifurcations from the primary finger states, although this time almost exclusively from states far beyond self-contact. Figure~\ref{fig:F_T0} shows a number of examples in the $(P,L)$ plane.
\begin{widetext}
\onecolumngrid\
\begin{center}\
\begin{figure}
    \centering
    \includegraphics[width=0.75\linewidth]{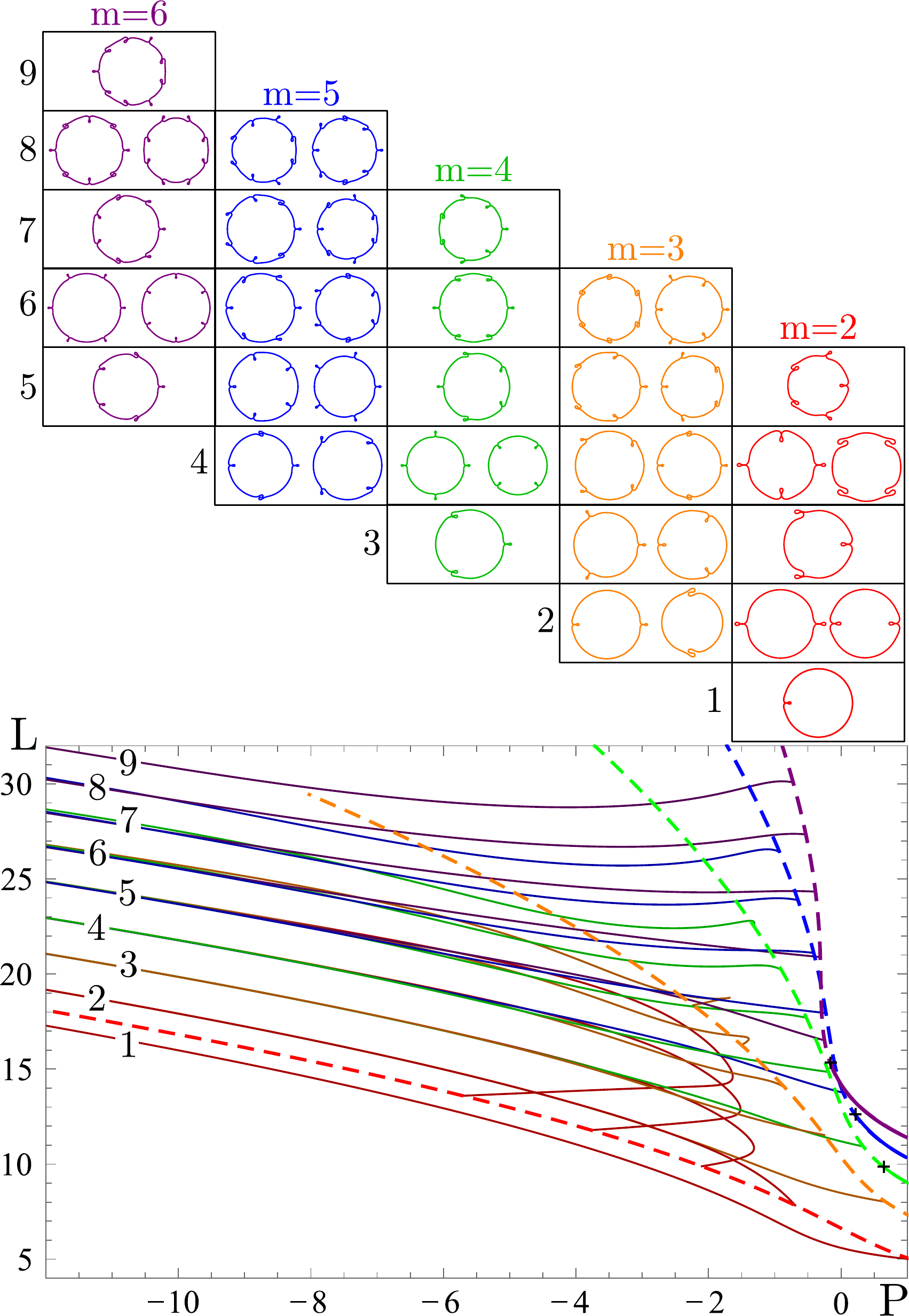}
    \caption{Bifurcation diagram depicting the $m=2,3,4,5,6$ finger states in thick lines (red, orange, green, blue, and purple, respectively), and the many fold states bifurcating from them (thin lines, corresponding color).  Self-contact for primary states is marked with a cross, and states beyond this point are dashed. Sample solutions (all rescaled to equal size for clarity) are arranged in ascending order according to the number of localized intrusions and protrusions on the $2\pi$ domain (profile labels correspond to branch labels in the bifurcation diagram). The number of distinct states created in each bifurcation grows in successive bifurcations as $P$ decreases ($L$ increases) and depends on the wave number $m$ of the underlying finger state. At large negative $P$ (large $L$) branches with the same number of intrusions and protrusions are almost degenerate since both contribute only slightly to the length, despite originating in general from different primary finger states as indicated by the color-coded profiles at the top panels. For example, there are four branches labeled 2, two of which originate from the $m=2$ finger branch (red) and two of which originate from the $m=3$ finger branch (orange). The branches originating in the same branch are strictly degenerate in this projection, although they reach self-contact at slighly different locations (not shown). Profiles originating in subsequent primary bifurcations are arranged vertically and correspond to $P\approx -20$.
    \label{fig:F_T0}}
\end{figure}
\end{center}\
\twocolumngrid\
\end{widetext}

We see that each primary finger branch (thick solid lines) yields a multitude of secondary branches of fold states with varying numbers of intrusions and protrusions, collectively referred to as folds. The first fold state to emerge from a primary finger state with wave number $m$ generates a state with $m-1$ folds, and each subsequent secondary bifurcation adds one additional fold. These folds can be intruding, protruding, or come in antisymmetric pairs (due to the symmetry constraint imposed by our boundary conditions).  There is always a special pair of intruding and protruding fold states which respect the symmetry of the finger state they bifurcate from.

As $P$ becomes more negative and the length of the domain increases, the profiles gradually deform from the self-intersecting, nonphysical states they bifurcate from into non-self-intersecting, physically realizable states.  Beyond the point of last self-contact, the profiles remain physical even as their length continues to increase.  Simultaneously, the details of the shape of the folds become less important, and the branches collapse to evenly spaced curves, each corresponding to a particular number of folds, as indicated on the left of the bifurcation diagram. Each of these asymptotic lines consists of multiple branches originating from different primary states. For example, there are four distinct 2-fold states, two of which are symmetric states with either two intrusions or two protrusions originating from the second primary bifurcation on the $m=2$ finger branch, and two additional states, one with one intrusion and one protrusion, and one with a pair of antisymmetric folds, originating from the first primary bifurcation on the $m=3$ finger branch. In the present projection, these pairs of branches lie on top of one another and one therefore sees only two branches, labeled 2. Likewise there are three branches labeled 3, originating in the third bifurcation on the $m=2$ branch, the second bifurcation on the $m=3$ branch and the first bifurcation on the $m=4$ branch, each with appropriate multiplicity (see profiles in row labeled 3). As in the case of the fold states for the constrained length (Fig.~\ref{fig:m9LS}), many fold states respect the symmetry of the primary branch they bifurcate from.  For example, the second pair of states bifurcating from $m=3$ features a solution with three intruding folds and a solution with three protruding folds, while $m=6$ bifurcates to a state with three pairs of antisymmetric folds and a solution with six alternating intrusions and protrusions with overall symmetry under rotations through $2\pi/3$.  The different states of alternating symmetric intrusions and protrusions as well as antisymmetric folds resemble the folded states of a floating planar elastic sheet under compression \cite{LeoEdgarsheet}, particularly when $L$ is large \footnote{The profiles in Fig.~4 of \cite{LeoEdgarsheet} are labeled in the opposite order to the branches in Fig.~3.}.

\section{Asymmetric states}\label{chiral}

\begin{figure}[h!]
    \centering
    \includegraphics[width=0.95\linewidth]{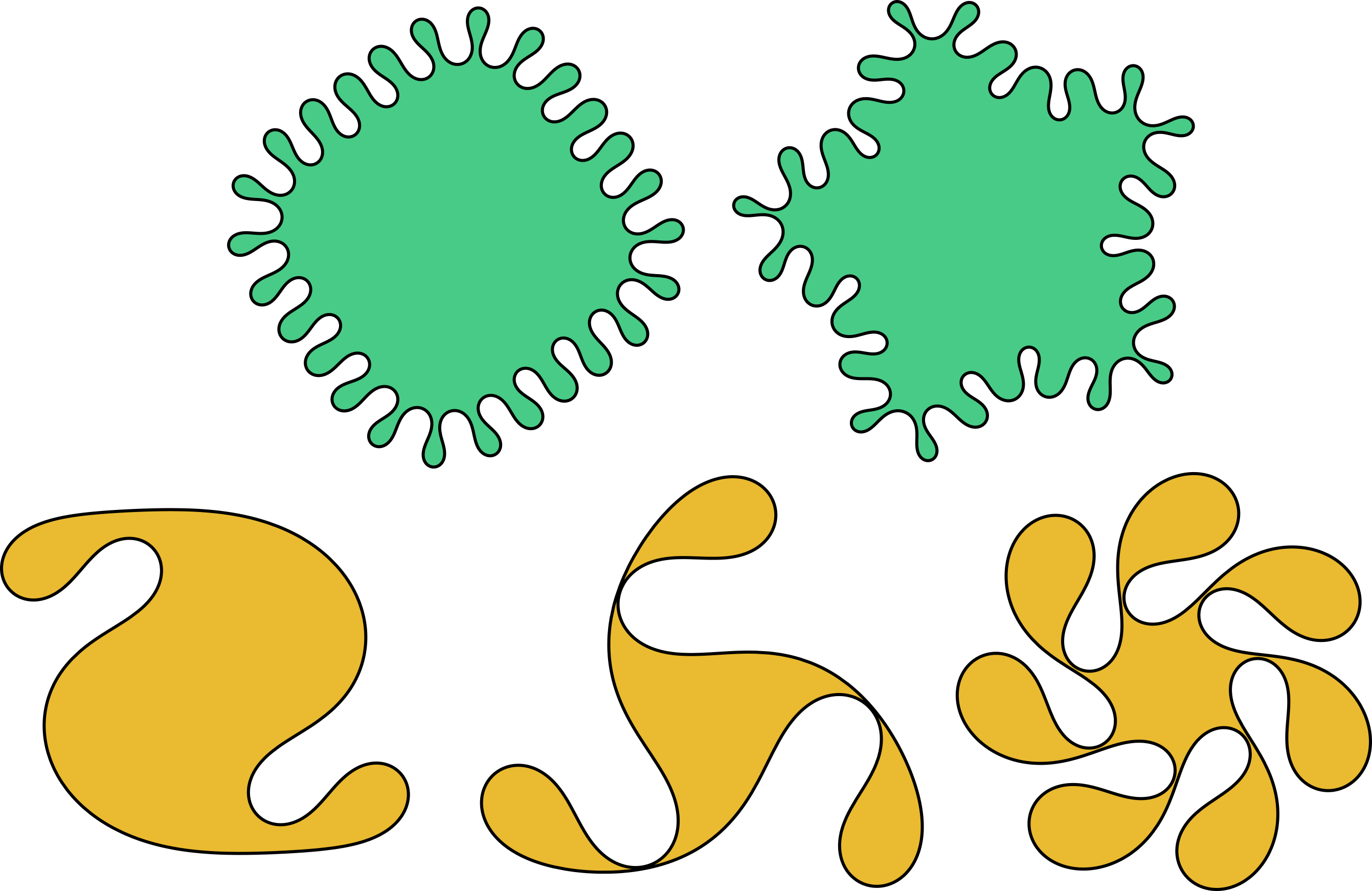}
    \caption{Asymmetric states computed on the full domain. Top left: $L=15.714$, $\ell^5=8232.56$; top right: $L=15.714$, $\ell^5=6257.5$; bottom left: $L=6.283$, $\ell^5=650$; bottom center: $L=25.649$, $\ell^5=0.916$; bottom right: $L=15.707$, $\ell^5=295$, all computed for $P=0$.}
    \label{fig:RK1}
\end{figure}

In the unconstrained case, many interesting steady-state finger profiles have been reported in the literature \cite{Carvalho2014_HS_elasticB}, but the continuation scheme summarized in Eqs.~(\ref{bcs}) rules out many of them since it can only generate solutions that are symmetric about the $x$ axis.  However, using an explicit Runge-Kutta method and numerical shooting, we are able to construct a variety of full-domain solutions which do not obey the previous symmetry restriction imposed by our continuation procedure. Figure~\ref{fig:RK1} shows examples that are similar in nature to those discovered in previous work \cite{Carvalho2014_HS_elasticB}. In the top row, we show mixed-mode solutions with no reflection symmetry, while the bottom row shows chiral versions of the localized states shown in Fig.~\ref{fig:m9LS}. Owing to the symmetry of Eq.~(\ref{eq:phiODE}) under reflection $(\phi,s)\to -(\phi,s)$, each left-handed solution is accompanied by an identical but right-handed solution.

\begin{figure}[h!]
    \centering
    \includegraphics[width=0.95\linewidth]{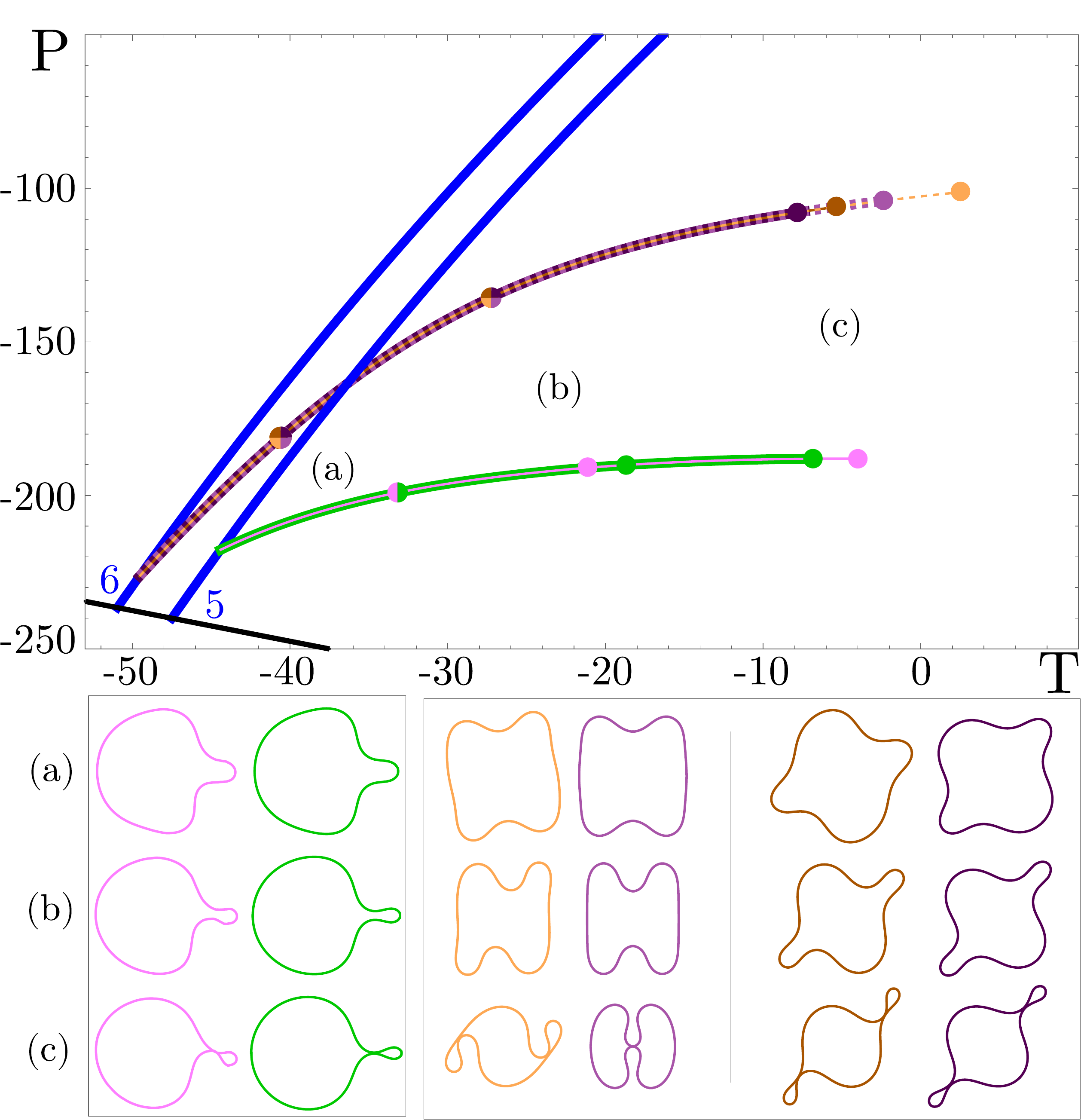}
    \caption{Full-domain numerical continuation for $\ell^5=576$ in the $(T,P)$ plane showing the first three symmetric and asymmetric fold states emerging simultaneously from secondary bifurcations of the $m=5$ and $m=6$ symmetric finger states (blue lines) which in turn bifurcate from the circle states (grey line). The branches of symmetric and asymmetric states coincide in this projection. Color-coded solution profiles corresponding to the locations indicated in the top panel are shown along the bottom.}
    \label{fig:s3bf}
\end{figure}

To determine the origin of these states in parameter space we implemented a full-circle extension of our numerical continuation in AUTO using periodic boundary conditions for $(\partial_s \phi, \partial_s^2 \phi, \partial_s^3 \phi, x, y)$ with Dirichlet boundary conditions for $\phi$ at $s=0,2\pi$ to pin the phase on the full domain.  The continuation of the first few primary and secondary branches of symmetric and asymmetric states is shown in Fig.~\ref{fig:s3bf}.


Figure~\ref{fig:s3bf} shows that asymmetric states appear via {\it secondary} bifurcations from the symmmetric finger states, a result consistent with the fact that all primary states in O(2)-symmetric steady state bifurcation problems are necessarily reflection-symmetric \cite{crawford}. It has previously been observed that the branches of symmetric intruding and protruding folds emerging from a common bifurcation point coincide in the $(T,P)$ plane \cite{foster_2021}. Figure~\ref{fig:s3bf} shows that in fact {\it all} secondary branches (intruding, protruding, symmetric, or chiral) emerging from a common bifurcation point coincide in the $(T,P)$ plane although they do not reach self-contact at the same location. We believe, but have been unable to check, that the multifold chiral states shown in Fig.~\ref{fig:RK1} likewise originate in secondary bifurcations from symmetric finger states.


\section{Conclusion}\label{conclusion}

We have studied the bifurcation properties of equilibrium states arising from a Rayleigh-Taylor-like instability of a higher density fluid confined within a lower density fluid in a rotating Hele-Shaw cell. The interface separating the two fluids is modeled as a thin elastic membrane, with the centrifugal force playing the role of effective outward gravity. The bending modulus of the interface introduces an intrinsic length scale into the problem that defines the scale of the resulting fingering instability. We examined two cases, one in which the interface was taken to be inextensible (the constrained case) and one in which the interface was permitted to proliferate freely (the unconstrained case). In both cases, we used numerical continuation to follow strongly nonlinear equilibrium finger states in parameter space through to the point of self-contact. We showed that, depending on parameters, these states may undergo secondary bifurcations leading to two types of secondary states, mixed modes and folds. The former form secondary connections between fingers with distinct wave numbers, while the latter form progressively more localized intrusions or protrusions as one follows each fold branch away from the secondary bifurcation that generates it.

The unconstrained or tension-free case is of particular interest. Here the primary finger states were found to be disconnected from the circle states, in contrast to the constrained case. However, as in the constrained case, the finger states exhibit instabilities to both mixed-mode states and to fold states. The latter take the form of symmetric intrusions or protrusions or antisymmetric folds that come in pairs to maintain the overall reflection symmetry imposed by our numerical continuation scheme. In particular, we demonstrated that the first bifurcation of a finger state with wave number $m$ generates a fold state with $m-1$ folds of various types. States with the same number of folds but different orientation or shape are created in bifurcations from other finger states with different but smaller wave number but these bifurcations are necessarily subsequent bifurcations and not the first. For example, the first bifurcation of the $m=3$ finger states creates a state with two folds, and so does the second bifurcation of the $m=2$ finger state (Fig.~\ref{fig:F_T0}). The folds localize away from these bifurcations as $P$ becomes more and more negative and the interface length grows, resulting in asymptotic degeneracy of all branches with the same number of folds, regardless of type and origin. This is a consequence of the fact that in this regime the folds take up an increasingly small fraction of the overall length $L$. The unconstrained system thus recapitulates similar behavior found earlier in a floating elastic sheet under compression \cite{diamant_witten2011,LeoEdgarsheet}, a system described by an equation similar to Eq.~(\ref{eq:phiODE}).

Moreover, the finger states can be either symmetric $m$-finger states, or break this symmetry, forming an $m$-finger chiral state of definite handedness. However, these chiral states appear via secondary bifurcations from an already existing finger state, at the same bifurcation as the symmetric folds, and cannot form in a primary bifurcation of the circle state. Together these results shed light on the origin and organization of the states reported previously in the unconstrained case with zero pressure difference across the interface \cite{Carvalho2014_HS_elasticB}.


Similar progressive localization of folds takes place in the constrained case as well, as the parameter $\ell$ or equivalently the rotation rate increases (Fig.~\ref{fig:LS2}). In both cases this is a consequence of the fact that the folds bifurcate {\it subcritically}, much as in the Swift-Hohenberg equation with a bistable nonlinearity \cite{knobloch2015}.


This work was supported in part by the National Science Foundation under grant DMS-$1908891$. We thank L. Gordillo and N. Verschueren for valuable discussions.

\newpage
\section*{References}{\bibliographystyle{unsrt}}

\providecommand{\noopsort}[1]{}\providecommand{\singleletter}[1]{#1}%

\end{document}